\documentclass[twocolumn,aps,prb,showpacs]{revtex4}
\usepackage{graphicx}
\usepackage{subfigure}

\begin{document}

\title{Re-$t_{2g}$-splitting-driven semiconductor gaps in ferrimagnetic double
perovskite Ca$_2$$M$ReO$_6$ ($M$=Cr,Fe) from first-principles}

\author{Sai Gong, San-Dong Guo, Peng Chen and Bang-Gui Liu}\email[Corresponding author:~]{bgliu@iphy.ac.cn}
\address{Beijing National Laboratory for Condensed Matter
Physics, Institute of Physics,  Chinese Academy of Sciences,
Beijing 100190, China}

\date{\today}

\begin{abstract}
Motivated by the observation of nonmetallic nature in double
perovskite Ca$_2$CrReO$_6$ and Ca$_2$FeReO$_6$ with high magnetic
Curie temperatures of 360 and 522 K, we systematically investigate
the structural, electronic, and magnetic properties of
Ca$_2$$M$ReO$_6$ ($M$=Cr,Fe) using the full-potential linear
augmented plane wave (FP-LAPW) method within the density functional
theory. Our full optimization confirms the stable ground-state
structure with $P2_1/n$ symmetry. The modified Becke-Johnson (mBJ)
exchange potential is used for investigating electronic structures.
Our mBJ calculation shows that they are both ferrimagnetic
semiconductors with semiconductor gaps of 0.38 eV and 0.05 eV,
respectively, in contrast with wrong metallic phases from the
generalized gradient approximation (GGA). The origin of
semiconductor gap is due to the further distortion of ReO$_6$
octahedra caused by John-Teller effect, which drives the three
partially-occupied Re $t_{2g}$ bands split into two fully-filled
bands and one empty band in the minority-spin channel. With the
spin-orbit coupling (SOC) taken into account, the Ca$_2$$M$ReO$_6$
($M$=Cr,Fe) shows high magneto-crystalline anisotropy (MCA) with the
magnetic easy axis along pseudocubic [010] direction, and the total
magnetic moments increase by 0.209$\mu_B$ and 0.258$\mu_B$ per
formula unit, respectively, due to the strong SOC effect on Re ion.
Although reducing to 0.31 and 0.03 eV, the semiconductor gaps remain
open in spite of the SOC broadening of the Re $t_{2g}$-related
bands. Therefore, our DFT investigation with mBJ has established the
correct ferrimagnetic semiconductor ground state for the double
perovskites Ca$_2$$M$ReO$_6$ ($M$=Cr,Fe). This mechanism, different
from that in double perovskite Sr$_2$CrOsO$_6$, can help understand
physical properties of other similar compounds.
\end{abstract}


\maketitle

\section{Introduction}

Because of their rich physics and high technological
potential\cite{9}, ordered double perovskite $A_2$$B$$B^\prime$O$_6$
($A$ = alkali, alkaline-earth or rare-earth ion; $B$ and $B^\prime$
= transition metals) have been extensively
studied\cite{1,2,3,4,5,6,7,8,10,11,14,16,add1,add2,add3,add4,52}.
For cubic or tetragonal double perovskite Sr$_2$$B$$B^\prime$O$_6$
($B$ = Cr or Fe, and $B^\prime$ = Mo, W, or Re), ferrimagnetic
metallic phase is usually formed because the fully occupied high
spin state Fe$^{3+}$ ($3d^5$) or Cr$^{3+}$ ($3d^3$) is
antiferromagnetically coupled with the partially filled $4d$ and
$5d$ transition-metal cations\cite{17,52}. Among them, half-metallic
Sr$_2$FeMoO$_6$, Sr$_2$FeReO$_6$, and Sr$_2$CrReO$_6$ have been
known as prospective spintronic materials beyond room
temperature\cite{2,3,10,19,52}. On the other hand, double perovskite
Sr$_2$CrOsO$_6$ is a robust ferrimagnetic insulator with the highest
magnetic Curie of 725 K, and its semiconductor gap has been shown to
originate from spin-exchange splitting of the Os 5$d$ $t_{2g}$
bands\cite{20,52,add5}. Very special are double perovskite
Ca$_2$FeReO$_6$ and Ca$_2$CrReO$_6$ whose ground-state phases are
ferrimagnetic insulators with monoclinic structure although there
are two electrons for their Re 5$d$ t$_{2g}$ triplet\cite{add1,52}.
They have high magnetic Curie temperatures of 522 and 360
K\cite{16,add1,52}. Their ferrimagnetic insulating phases have been
established experimentally\cite{19,16,38,add1,52,33}, but their
non-metallic electronic properties have not been elucidated yet and
their structure-property relationship still needs to be understood,
especially in the case of the Ca$_2$CrReO$_6$.

Here, we investigate the structural, electronic and magnetic
properties of the Ca$_2$CrReO$_6$ and Ca$_2$FeReO$_6$ through
density functional theory calculations in order to reveal the origin
of their special electronic structures, especially their
semiconductor gaps at low temperature. To accurately calculate the
semiconductor gaps, we use Tran and Blaha¡¯s modified Becke and
Johnson (mBJ) approach for the exchange potential \cite{40} to
investigate their electronic structures, because its excellent
accuracy has been proved for numerous insulators, semiconductors,
and transition-metal oxides\cite{40,41,43,44}. Our calculations and
analyses show that the Ca$_2$CrReO$_6$ and Ca$_2$FeReO$_6$ still are
ferrimagnetic semiconductors even with the spin-orbit effect taken
into account, and the semiconductor gaps are formed between the
full-filled $d_{xy}$+$d_{xz}$ doublet and the empty $d_{yz}$ singlet
split from the partially-occupied Re 5d $t_{2g}$ triplet around the
Fermi level due to the low symmetry in the monoclinic structure. We
also explore other properties of the two Ca-based double perovskite
compounds in comparison with others similar. More detailed results
will be presented in the following.

The rest of the paper is organized as follows. We shall describe our
computational details in the next section. In Sec. III we shall
present our optimized ground-state structures for the two compounds.
In Sec. IV we shall present our spin-dependent density of states,
band structures, and electron density distributions and perform
further analyses concerned. In Sec. V we shall present our
calculated results with the spin-orbit effect taken into account,
including their magneto-crystalline anisotropic energies, spin and
orbital moments along the easy axis, and the
spin-orbit-effect-modified semiconductor gaps. Finally, we shall
give our conclusion in Sec. VI.

\section{Computational details}

We use the full-potential linear augmented
plane wave (FP-LAPW) method within the density functional theory (DFT),\cite{46,47}
as implemented in the WIEN2k package.\cite{48}
We take GGA exchange-correlation functional to do structure optimization and preliminary study\cite{39},
and then use mBJ exchange potential to do electronic structure calculations. The scalar relativistic
approximation is used for valence states, with the spin-orbit coupling (SOC)
is taken into account, whereas the radial Dirac equation is self-consistently solved for the core electrons\cite{49,50,51}. The magnetization is chosen to be along all nonequivalent directions for the monoclinic
structure  when we investigate the magneto-crystalline anisotropy.
The muffin-tin radii of the Ca, Cr, Fe, Re, and O atoms are set to be
2.20, 1.96, 2.03, 1.96, and 1.71 bohr, respectively. We make
harmonic expansion up to $l_{max}$ = 10 in the muffin-tin spheres, and set $R_{mt}\times
K_{max}$ = 8.0. We use 1000 k-points in whole Brillouin zone (234 k-points
in the reduced wedge). For testing the accuracy, we also use 2000 k-points to do
the self-consistent calculations. The total energy difference is proved to be less than
1 meV. Therefore, our choice of 1000 k-points is enough for the whole calculation.
The self-consistent calculations are considered to
be converged only when the integration of absolute charge-density
difference per formula unit between the successive loops is less
than $0.0001|e|$, where $e$ is the electron charge.

\section{Structure optimization}

The structure of Ca$_2$$M$ReO$_6$ ($M$=Cr or Fe) has been reported to be
monoclinic structure with $P2_1/n$ symmetry (space group \#14) at room temperature,\cite{16,52}
which is consistent with the prediction of the empirical tolerance factor $f$.\cite{33,52}
The monoclinic structure is fairly distorted from cubic double perovskite
due to the small size of Ca$^{2+}$ cation, which forces the $M$O$_6$ and ReO$_6$
octahedra to tilt and rotate in order to optimize the Ca-O bond lengths.
The crystal structure for Ca$_2$$M$ReO$_6$ ($M$=Cr or Fe) is demonstrated in Fig. 1.
In order to investigate the origin of the nonmetallicity in their ground-state phases, we
optimize fully their geometric structures and internal atomic positions
by combining total energy and force optimizations.
We have considered a larger unit cell of 2 f.u. including 20
atoms to relax the structure. The optimized lattice parameters are listed
in Table \ref{table1}, with experimental data included for comparison.
The ground state phase is confirmed to be the $P2_1/n$ structure for both of the compounds, with lattice
constants expanded slightly with respect to experimental ones. This deviation is due to the special property of the GGA functional.
However, the tilt angles $\beta$ of Ca$_2$$M$ReO$_6$ ($M$=Cr and Fe) decrease by
0.22$^\circ$ and 0.27$^\circ$ with respect to experimental values, respectively,
which reflects the large distortion at low temperature.

\begin{figure}[htb]
\includegraphics[width=6cm]{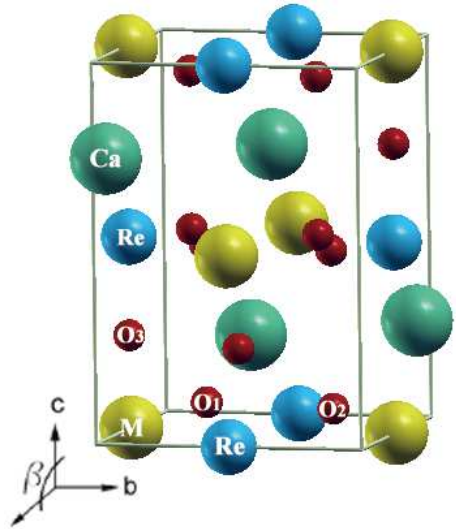}
\caption{(color online) The $P2_1/n$ (\#14) crystal structure of
double perovskites Ca$_2$$M$ReO$_6$, with three nonequivalent
O atoms labeled. Ca, $M$, Re, and O atoms are denoted by colored balls
with different sizes from the biggest to the smallest, where $M$ denotes Cr or Fe atom.
The $M$ and Re atoms form a NaCl-like lattice.}\label{fig1}
\end{figure}

\begin{table}[htb]
\caption{Optimized lattice constants and tilt angle $\beta$
of Ca$_2$$M$ReO$_6$ ($M$=Cr,Fe) with space group $P2_1/n$ (\#14), compared with experimental results\cite{16}}\label{table1}
\begin{ruledtabular}
\begin{tabular}{ccccc}
 Lattice &\multicolumn{2}{c}{Ca$_2$CrReO$_6$} & \multicolumn{2}{c}{Ca$_2$FeReO$_6$}  \\
\cline{2-5}
 parameters & opt. & exp. & opt. & exp. \\
\hline
 a (\AA) & 5.392 & 5.388 & 5.417 & 5.401  \\
 b (\AA) & 5.524 & 5.460 & 5.609 & 5.525  \\
 c (\AA) & 7.680 & 7.660 & 7.733 & 7.684  \\
\hline
 $\beta$ ($^\circ$) & 89.74 & 89.96 & 89.80 & 90.07  \\
 \end{tabular}
 \end{ruledtabular}
\end{table}

We present in Table \ref{table2} optimized bond lengths and bond angles
of Ca$_2$$M$ReO$_6$ ($M$=Cr,Fe) with $P2_1/n$ structure. For comparison, we
also present the $I4/m$ structure for Ca$_2$CrReO$_6$. It's shown that the three Cr-O bond lengths are slightly larger
than Re-O ones in Ca$_2$CrReO$_6$, while the FeO$_6$ octahedra are significantly more
expanded than ReO$_6$ octahedra in Ca$_2$FeReO$_6$. This observation is consistent with the ionic
size sequence of Re$^{5+}$ $<$ Cr$^{3+}$ $<$ Fe$^{3+}$. The bond angles in ReO$_6$
octahedra all deviate from ideal values of 90$^\circ$, so do the angles of $M$-O-Re
from 180$^\circ$. Considering that a large number of double perovskite compounds with half-metallicity are in tetragonal structure with $I4/m$ symmetry (space group \#87), and in order to clarify the relationship between the electronic property
and lattice structure, we also present the structure parameters of the $I4/m$ structure for Ca$_2$CrReO$_6$ in Table \ref{table2}. There are actually two nonequivalent kinds of O atoms in $I4/m$ structure. The O$_1$ and O$_2$ atoms are
equal to each other and in the same $xy$-plane. The atom O$_3$ sits along the $z$-axis with
Cr or Re atoms in between. The bond lengths of Cr-O are larger than those of Re-O
due to the larger ionic size of Cr$^{3+}$ versus Re$^{5+}$. All angles in ReO$_6$
octahedra remain to be 90$^\circ$, while the Cr-O$_{1,2}$-Re bond angles reduce significantly
from 180$^\circ$, which makes the Cr-O$_{1,2}$ and Re-O$_{1,2}$ lengths
much larger than the Cr-O$_3$ and Re-O$_3$ bonds, respectively. Our calculated total
energy of Ca$_2$CrReO$_6$ in $I4/m$ structure are higher than that in
$P2_1/n$ structure by 392 meV per formula unit, indicating the $P2_1/n$ structure is more stable
for Ca$_2$CrReO$_6$.

\begin{table}[htb]
\caption{Optimized bond lengths and angles of
Ca$_2$$M$ReO$_6$ ($M$=Cr, Fe) with space group $P2_1/n$ (\#14),
with those of Ca$_2$CrReO$_6$ with $I4/m$ (\#87) structure for comparison. O$_1$ and O$_2$ are
equivalent to each other in $I4/m$ structure.}\label{table2}
\begin{ruledtabular}
\begin{tabular}{ccccc}
& $M$ & Cr(\#14) & Cr(\#87) & Fe(\#14) \\
\hline
Bond length & M-O$_1$ & 1.975 & 1.984 & 2.061 \\
            & M-O$_2$ & 1.978 & 1.984 & 2.046 \\
            & M-O$_3$ & 1.971 & 1.939 & 2.036  \\
            & Re-O$_1$ & 1.974 & 1.977 & 1.944 \\
            & Re-O$_2$ & 1.968 & 1.977 & 1.948 \\
            & Re-O$_3$ & 1.964 & 1.922 & 1.941 \\
\hline
Bond angle & O$_1$-Re-O$_2$ & 90.88 & 90 & 90.30 \\
           & O$_2$-Re-O$_3$ & 89.43 & 90 & 89.28 \\
           & O$_1$-Re-O$_3$ & 89.81 & 90 & 89.22 \\
\cline{2-5}
           & M-O$_1$-Re & 152.56 & 154.63 & 149.34 \\
           & M-O$_2$-Re & 152.85 & 154.63 & 150.59 \\
           & M-O$_3$-Re & 153.46 & 180    & 150.06 \\
 \end{tabular}
 \end{ruledtabular}
\end{table}

\section{Electronic structures}

\subsection{Density of states and energy bands}

\begin{figure}[htb]
\centerline{\includegraphics[width=7cm]{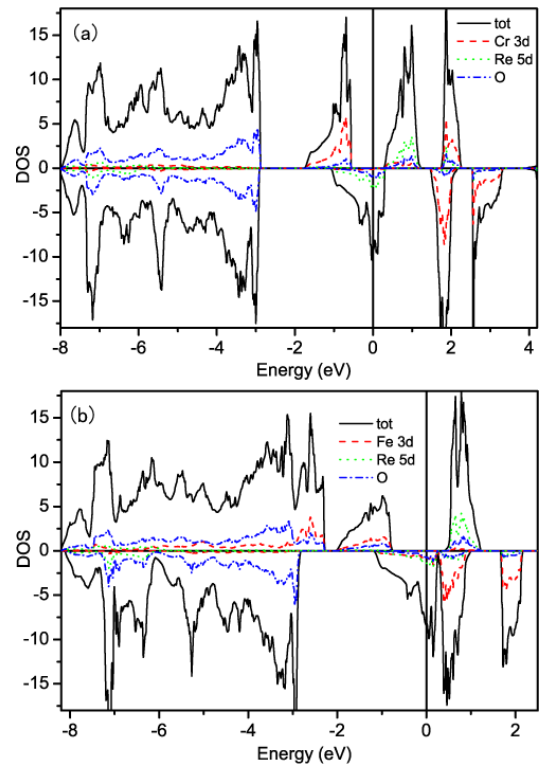}}
\caption{(color online) The spin-resolved total (tot) and partial (Cr/Fe,Re,O) density of states (DOSs)
of Ca$_2$CrReO$_6$ (a) and Ca$_2$FeReO$_6$ (b) in $P2_1/n$ structure from
GGA calculation.}\label{dos GGA}
\end{figure}

From now on, we investigate the electronic structures of the optimized Ca$_2$$M$ReO$_6$ ($M$=Cr,Fe).
At first, we use the popular GGA functional to calculate the density of states (DOS).
The spin-resolved DOSs are presented in Fig. 2. For the Ca$_2$CrReO$_6$, the electronic energy
bands between -8.0 eV and -3.0 eV are dominated by
O $2p$ states. The Fermi energy falls in an energy gap of about 1.0 eV in the majority spin channel, between the
fully filled Cr $t_{2g}$ and empty Re $t_{2g}$ bands. As for the
Ca$_2$FeReO$_6$, the triplet Fe $t_{2g}$ states in the majority spin channel move to
the lower energy between -8.2 eV and -2.2 eV, with a strong mixture of O $2p$ states.
The Fermi energy is in the majority-spin gap of 1.4 eV between Fe e$_g$ and
Re $t_{2g}$ bands. In contrast, for the minority spin channel, the Fermi level lies in the
partially filled $t_{2g}$ bands of hybridized $M$, Re, and O $2p$ states in the Ca$_2$$M$ReO$_6$ ($M$=Cr,Fe). Thus, the GGA calculation produces a half-metallic ferrimagnet.
This is contradictory with the reported experimental results\cite{16,52}
and can be attributed to the false GGA description of Re $t_{2g}$ nature around the Fermi level in the
monoclinic structure. For the sake of accurate calculation for Re $t_{2g}$ state,
we need to use improved exchange potential to investigate electronic structures
of the Ca$_2$$M$ReO$_6$. The modified Becke-Johnson (mBJ) potential is a good choice because it is excellent in describing the hybrid
transition-metal ions.\cite{41,53}

\begin{figure}[htb]
\centerline{\includegraphics[width=7.8cm]{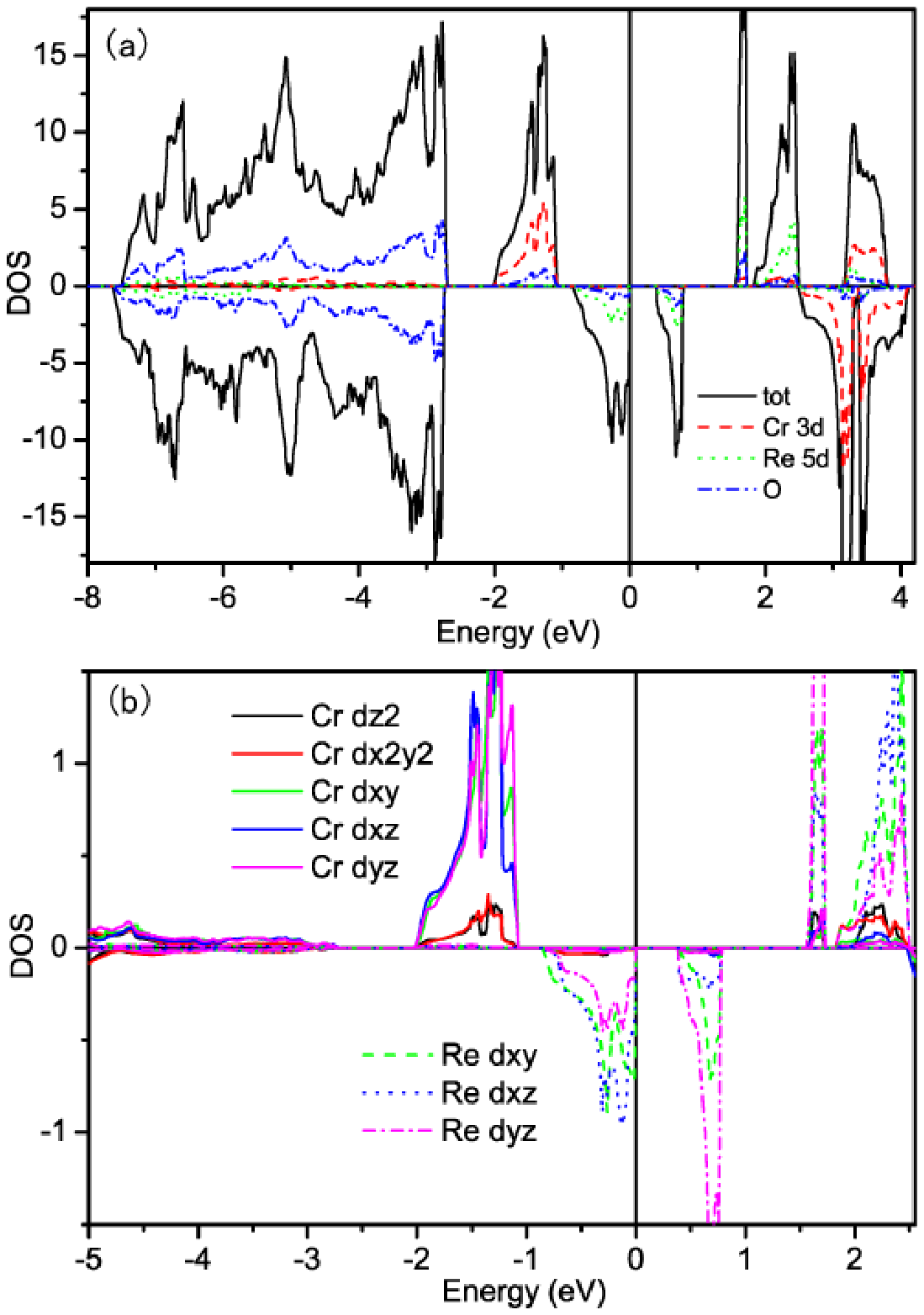}}
\vspace{0.1cm}
\centerline{\includegraphics[width=8cm]{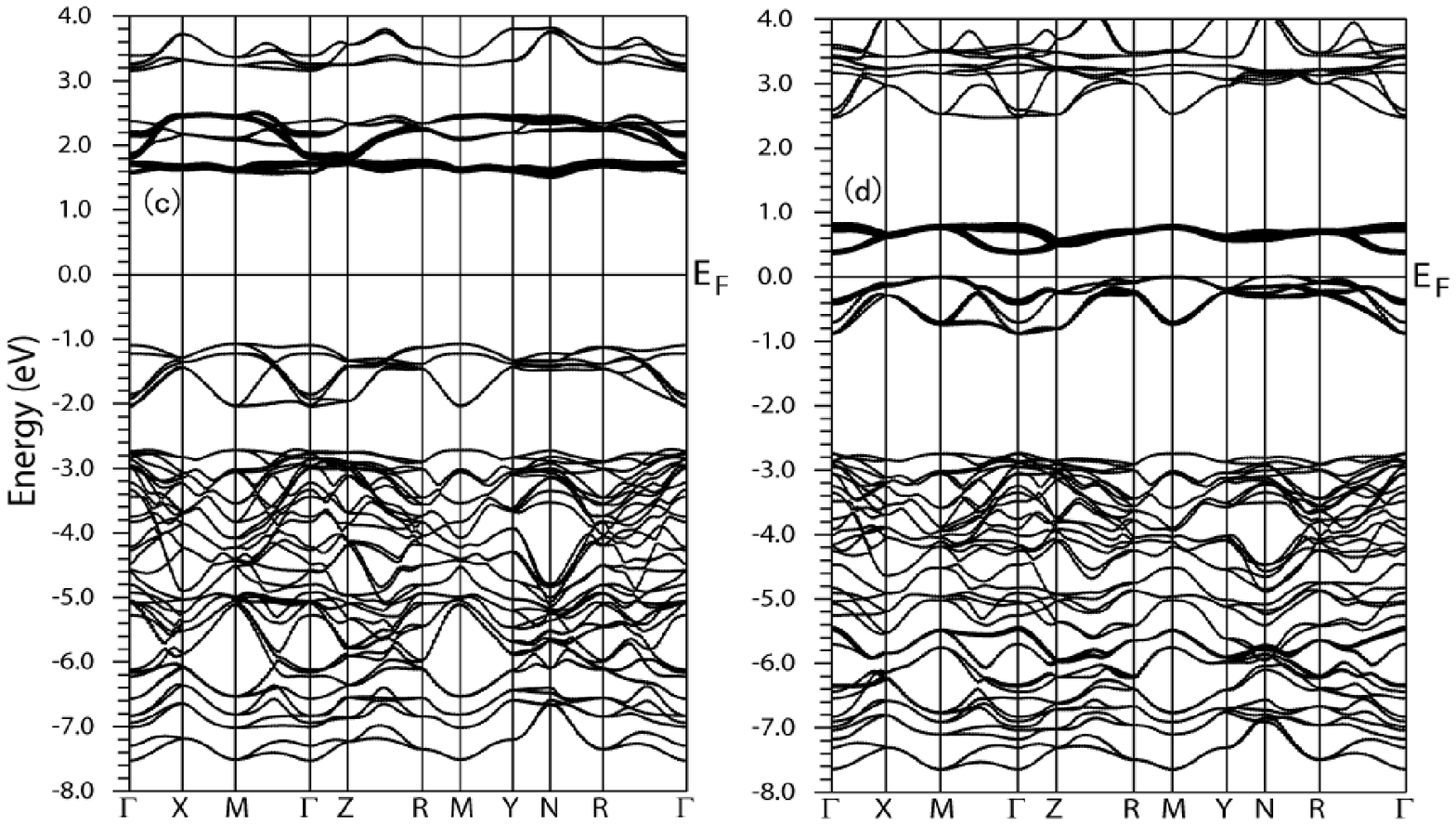}}
\caption{(color online) Electronic structure of Ca$_2$CrReO$_6$ in $P2_1/n$ structure calculated with mBJ: spin-resolved total (tot) and partial (Cr,Re,O) DOSs (a); amplified $d$ states split DOSs of Cr and Re around the Fermi level (b);
and majority-spin (c) and minority-spin (d) bands. }\label{fig2}
\end{figure}

\begin{figure}[htb]
\centerline{\includegraphics[width=7.8cm]{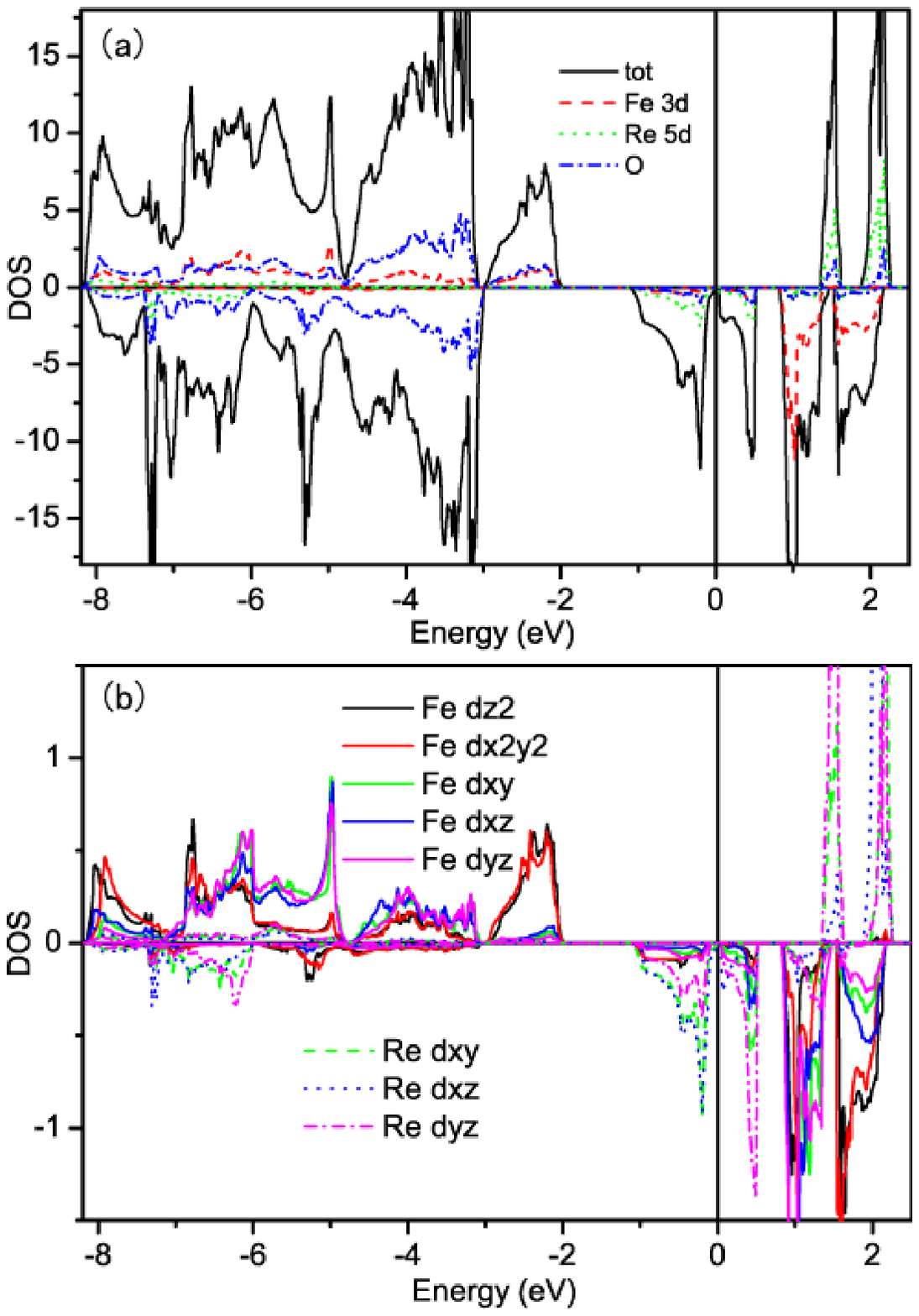}}
\vspace{0.1cm}
\centerline{\includegraphics[width=8cm]{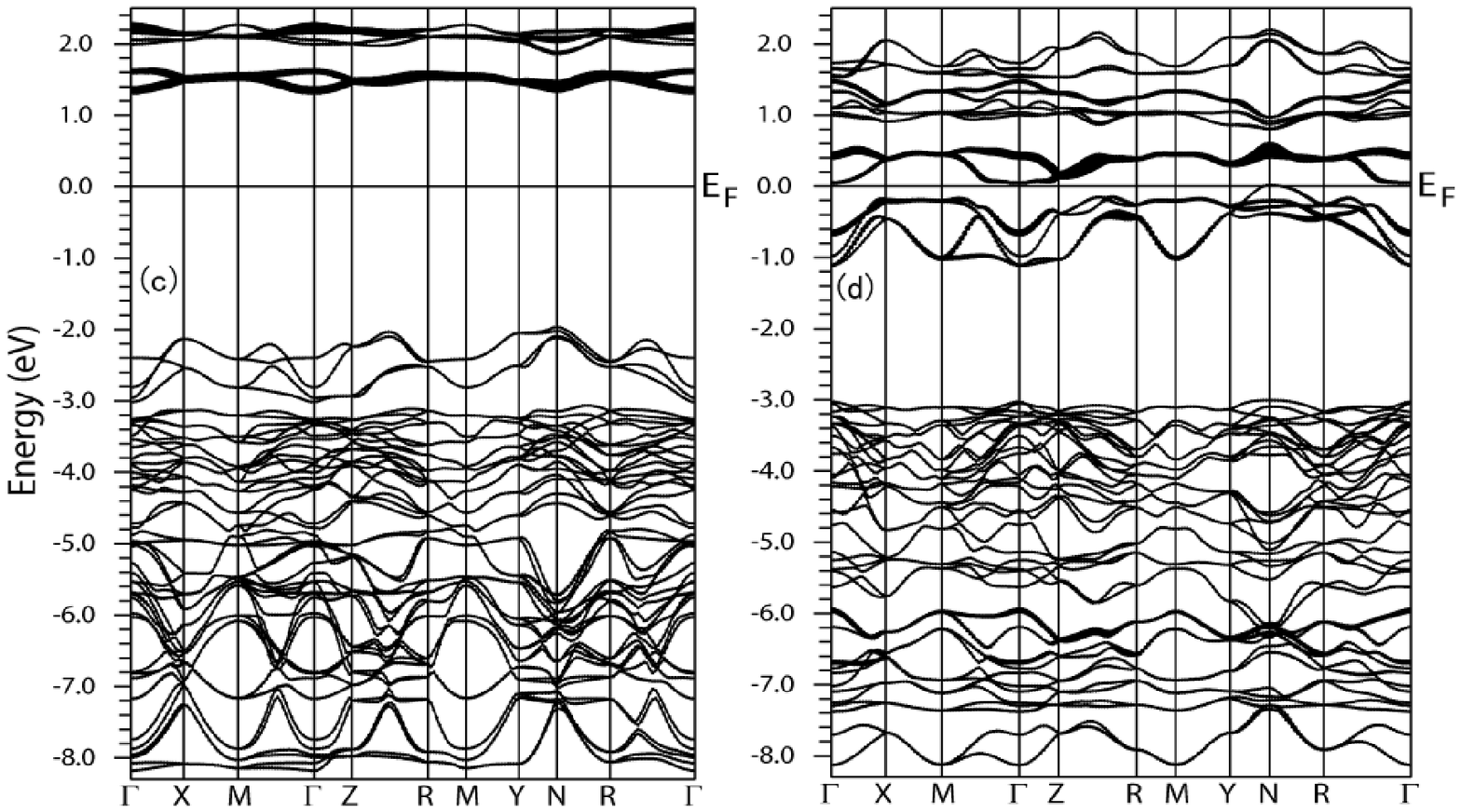}}
\caption{(color online) Electronic structure of Ca$_2$FeReO$_6$ in $P2_1/n$ structure calculated with mBJ: spin-resolved total (tot) and partial (Fe,Re,O) DOSs (a); amplified $d$ states split DOSs of Fe and Re around the Fermi level (b);
and majority-spin (c) and minority-spin (d) bands. }\label{fig3}
\end{figure}

We present the spin-resolved DOSs and energy bands of the Ca$_2$CrReO$_6$ calculated with mBJ in Fig. 3.
It is clear that there is a semiconductor gap open at the Fermi level, which is in good agrement with experimental results\cite{16}.
Moreover, the occupied Cr $t_{2g}$ and unoccupied Cr $e_g$ and Re $t_{2g}$
bands in the majority spin channel are pushed substantially downwards and upwards, respectively,
which consequently enhances the majority-spin gap (G$_{\rm mag}$) to 2.5 eV. In the minority
spin channel, the triplet Re $t_{2g}$ states around the Fermi level further split into a doublet $d_{xy}$+$d_{xz}$
and a singlet $d_{yz}$, with two electrons fully occupying the doublet state.
This produces a semiconductor gap of 0.38 eV, as shown in Fig. 3(a). The detailed orbital-resolved DOSs around the Fermi level
are presented in Fig. 3(b). Both of the unoccupied Cr $t_{2g}$ and $e_g$
in the minority spin channel are pushed upwards substantially, with a little overlap
between them, which enlarges the spin exchange splitting energy of Cr $t_{2g}$ to
4.7 eV. Fig. 3(c) and (d) show the energy bands of the Ca$_2$CrReO$_6$
in majority-spin and minority-spin channels, respectively. The thicker a line is, the more the Re $t_{2g}$ weight is.
There are 72 bands (36 majority spin and 36 minority spin) in the energy window
from -7.8 eV to -2.6 eV, because we consider 2 unit cell in our calculation.
The energy bands between -2.0 eV and -1.0 eV consist of 6 Cr $t_{2g}$ bands in the
majority spin channel. There are 4 occupied bands of Re $t_{2g}$, including d$_{xy}$
and d$_{xz}$ just below the Fermi level and 2 bands of d$_{yz}$ above the Fermi level.
It can be clearly seen that the top of the valence band and the bottom of the conduction
band are located in N and $\Gamma$ points, respectively, resulting in an
indirect band gap for the Ca$_2$CrReO$_6$.

In Fig. 4 we present the spin-resolved DOSs and energy bands of the optimized Ca$_2$FeReO$_6$
calculated with mBJ. The distinguished feature of the DOSs
is that compared to GGA results, the Fe $t_{2g}$ states move to lower energy
in the majority spin channel, and the gaps between Fe $t_{2g}$ and $e_g$ states almost
vanish in both of spin channels. This can be attributed to the enhanced spin exchange effect due to mBJ functional.
The semiconductor gap of 0.05eV is also observed, owing to the same
reason as in the Ca$_2$CrReO$_6$. However, the Re $t_{2g}$ splitting in the minority
spin channel is much smaller than that in the majority spin channel, in contrast
to the Ca$_2$CrReO$_6$. This result is consistent with the fact
that the resistivity of the Ca$_2$FeReO$_6$ at low temperature is almost two order
of magnitude lower than that of Ca$_2$CrReO$_6$, because the electron
inter-band transition takes place easily in the Ca$_2$FeReO$_6$ compound. In the band structures (c) and (d)
of the Ca$_2$FeReO$_6$, the Fe $t_{2g}$ bands are pushed down to
between -8.0 eV and -3.0 eV in the majority spin channel, in contrast to the minority-spin one.
the energy window from -3.0 eV to -2.0 eV
consists of 4 bands of Fe $e_g$. The energy band distribution around the Fermi
level in the minority spin channel is similar to that of the Ca$_2$CrReO$_6$. The top
of valence bands and the bottom of the conduction bands are located in N and $\Gamma$
points, respectively. This implies that the semiconductor gap of the Ca$_2$FeReO$_6$ is indirect, same as that of the Ca$_2$CrReO$_6$.

\subsection{Electron density distributions}

\begin{figure*}[htb]
\begin{minipage}{0.7\textwidth}
  \includegraphics[width=11cm]{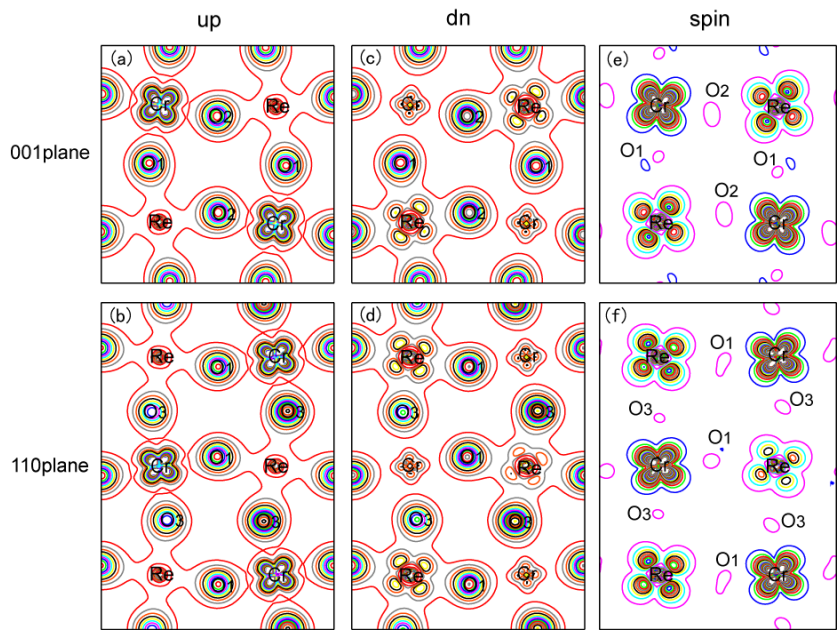}
 \end{minipage}%
 \begin{minipage}{0.3\textwidth}
 \caption{(color online) Valence electron charge [up, (a) and (b); dn, (c) and (d)] and spin [(e),(f)] density distributions, within the energy window from -8.0 eV to the Fermi level, of Ca$_2$CrReO$_6$ projected to the (001) and (110) planes calculated with mBJ.
The contours in (a)-(d) are from 0.005 to 0.5$e$/a.u.$^3$ with an increment of 0.025$e$/a.u.$^3$, and those in (e) and (f) from
-0.5 to 0 $e$/a.u.$^3$ for Re sites and 0 to 0.5$e$/a.u.$^3$ for Cr sites with an increment of 0.013$e$/a.u.$^3$.}\label{CaCrReO ED}
 \end{minipage}
\end{figure*}

\begin{figure*}[htb]
\begin{minipage}{0.7\textwidth}
  \includegraphics[width=11cm]{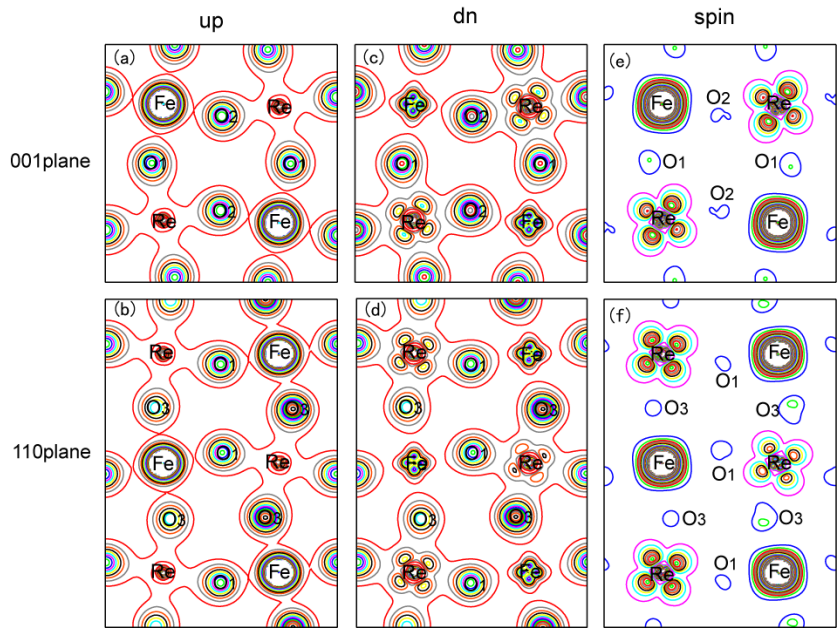}
 \end{minipage}%
 \begin{minipage}{0.3\textwidth}
 \caption{(color online) Valence electron charge [up, (a) and (b); dn, (c) and (d)] and spin [(e),(f)] density distributions, within the energy window from -8.2 eV to the Fermi level, of Ca$_2$FeReO$_6$ projected to the (001) and (110) planes calculated with mBJ.
The contours in (a)-(d) are from 0.005 to 0.5$e$/a.u.$^3$ with an increment of 0.025$e$/a.u.$^3$, and those in (e) and (f) from
-0.5 to 0 $e$/a.u.$^3$ for Re sites and 0 to 0.5$e$/a.u.$^3$ for Fe sites with an increment of 0.013$e$/a.u.$^3$.}\label{CaFeReO ED}
 \end{minipage}
\end{figure*}

The energy-resolved charge and spin density distributions are very important to
explore the bonding and magnetic properties. We present in Fig. 5 the valence
charge (including up and down) and spin density distributions of the Ca$_2$CrReO$_6$,
with all the contribution from -8.0 eV to the Fermi level, calculated with mBJ. The upper three
panels are for the (001) plane of the structure shown in Fig. 1, the
lower three ones are for the perpendicular plane, being equivalent to the (110) plane, including Cr and
Re ions. In the spin-up channel, the charge density
distributions at Cr sites look like a quatrefoil, which reflects the fully occupied
$t_{2g}$ characteristic between -2 eV to -1 eV, whereas the charge density around
Re is fairly small. In the spin-down channel,
there is much electron density around Re, in consistence
with the partially filled Re $t_{2g}$ state from -1 eV to the Fermi level.
The small charge density at Cr sits should result from the hybridization between Re and Cr $t_{2g}$ states.
In both of the spin channels, the oxygen
atoms with high electron affinities attract the electrons from Cr and Re atoms
to form nearly closed O $2p$ shells with spherically distributed charge densities.
It can be seen in the charge density contours that the bonds between Cr and nearest O
are almost ionic with respect to the Re-O bonds with covalent characteristic, which
is in accordance with the longer bond lengths of Cr-O than Re-O ones, as described
in Table \ref{table2}. The charge distributions also show that there exists
no direct interaction between two nearest Cr-Cr or Re-Re pairs. The spin density
distribution of the Ca$_2$CrReO$_6$ in Fig. 5 (e) and (f) demonstrates that the spin moments
of Cr and Re are mainly localized at the ionic sites. The different colors of
outermost lines from Cr and Re indicate the antiferromagnetic coupling between
the two transition-metal moments in the double perovskite Ca$_2$CrReO$_6$.
The some deformed quatrefoils of Cr and Re ions are ascribed to the
distortion of O octahedra. It is worth notice that different
density contours between Re and Cr sites are due to the more closed shells in the inner part
of heavier Re ion compared to Cr.

As for the Ca$_2$FeReO$_6$, we illustrate the corresponding charge and spin
density distributions calculated with mBJ in Fig. 6. It can be seen that charge distributions at the Fe site
are nearly spherical because of nearly half-filled Fe $3d$ orbitals, which is different
from the quatrefoils shape of partially occupied Cr $3d$ orbitals in the Ca$_2$CrReO$_6$.
The Fe $3d$ electrons that are more than half full move to oxygen sites for stabilizing
the ground state, leaving the highly ionized Fe atoms, as shown in Fig. 6.
The shape of charge density around the Re site is similar to that in the Ca$_2$CrReO$_6$
due to the same valence states of Re$^{5+}$ ions in the two compounds. Most of the
Re $5d$ electrons with larger orbitals spread out to O
$2p$ states, forming the Re-O covalent bonds.
Furthermore, the hybridizations are still along the Fe-O-Re-O-Fe
chain, and no direct interaction between Fe-Fe and Re-Re
pairs are found. The spin moments of Fe $3d$ and Re $5d$ states are
mainly localized and coupled antiferromagnetically.

\subsection{Further analyses}

The spin exchange splitting ($\Delta$ex) and crystal field splitting ($\Delta$cf)
of Cr and Fe, the spin exchange splitting $\Delta$ex of Re ion, and the band gaps across the Fermi level
in both majority-spin
(G$_{maj}$) and minority-spin (G$_{min}$)  channels calculated with GGA and mBJ
are summarized in Table \ref{table3}. Both $\Delta$ex and $\Delta$cf of
transition-metals are significantly enhanced by mBJ calculation. As a result,
the gaps in the majority spin channel are enlarged by 1.5 eV and 1.9 eV for the Ca$_2$$M$ReO$_6$ ($M$=Cr,Fe),
respectively. The semiconductor gaps are equivalent to 0.38 eV and
0.05 eV, respectively, in contrast to the wrong results from GGA. This implies
that the mBJ functional is excellent in describing the hybrid correlated
transition-metal ions in Ca$_2$$M$ReO$_6$ ($M$=Cr,Fe) compounds.

\begin{table}[htb]
\caption{Spin exchange splitting ($\Delta$ex) and crystal field splitting ($\Delta$cf)
of Cr and Fe, spin exchange splitting $\Delta$ex of Re ion, the band gaps across the Fermi level in the majority-spin channel
(G$_{\rm maj}$) and the minority-spin channel (G$_{\rm min}$) of Ca$_2$$M$ReO$_6$ ($M$=Cr, Fe)
calculated with GGA and mBJ.}\label{table3}
\begin{ruledtabular}
\begin{tabular}{ccccccc}
 $M$ & scheme & $\Delta$ex($M$) & $\Delta$cf($M$) & $\Delta$ex(Re) & G$_{\rm maj}$ & G$_{\rm min}$  \\
\hline
 Cr & GGA (eV) & 2.9 & 3.0 & 0.6 & 1.0 & 0  \\
      & mBJ (eV) & 5.3 & 4.7 & 2.0 & 2.5 & 0.38  \\ \hline

 Fe & GGA (eV) & 3.0 & 1.0 & 0.8 & 1.4 & 0  \\
      & mBJ (eV) & 4.5 & 1.5 & 1.8 & 3.3 & 0.05  \\
 \end{tabular}
 \end{ruledtabular}
\end{table}

For investigating the intrinsic reason of semiconductor nature in the monoclinic
Ca$_2$$M$ReO$_6$ ($M$=Cr,Fe), we calculate the DOSs of the Ca$_2$CrReO$_6$ in $I4/m$
structure with both GGA and mBJ functionals. The GGA
produces a half-metallicity, which is similar to the GGA result of the Ca$_2$CrReO$_6$
with $P2_1/n$ symmetry, and however, the metallic property is not changed by using mBJ potential,
although the $\Delta$ex and $\Delta$cf of Cr and Re are much enlarged.
This comparison indicates that the semiconductor nature of the Ca$_2$$M$ReO$_6$ has
intimate relationship with the crystal structure. In $I4/m$ structure, the bond
angles of ReO$_6$ octahedra are ideal 90$^\circ$. The Re $5d$ states are
divided into triply degenerate $t_{2g}$ states and doubly degenerate $e_g$ states.
Thus, the Fermi level across the partially filled Re $t_{2g}$ in the minority-spin
channel. As for $P2_1/n$ structure, the ReO$_6$ octahedra undergoes tilting and
rotation, making the bond angles all deviate from 90$^\circ$, as
listed in Table \ref{table2}. Therefore, the triplet Re $t_{2g}$ states are further
split into a doublet $d_{xy}$+$d_{xz}$ and a singlet $d_{yz}$ due to the Jahn-Teller
distortion, producing a crystalline field splitting band gap, in contrast to the
former Mott-Hubbard gap.\cite{16,19,38} This fact is consistent with the quatrefoil pattern of the spin density distribution at Re sites, as shown in Fig. 5(e) and (f) and Fig. 6 (e) and (f). The larger gap obtained in the Ca$_2$CrReO$_6$
compound can be attributed to the stronger distortion of ReO$_6$ octahedron than in the Ca$_2$FeReO$_6$.

\section{Spin-orbit coupling effect}

To investigate the effect of spin-orbit coupling on the optimized Ca$_2$$M$ReO$_6$ ($M$=Cr,Fe),
the magnetization is chosen to be approximately along [100], [010], [001], [110],
[101], [011] and [111] directions for the $P2_1/n$ monoclinic structure. The calculated
total energies with GGA+SOC method presented in Table \ref{table4} indicate high magneto-crystalline
anisotropy (MAC) in the Ca$_2$$M$ReO$_6$ compounds. The most stable magnetic orientation
are both along [010] direction, equivalent to the $b$-axis perpendicular to the $ac$ plane in the $P2_1/n$ structure. These are consistent with experimental results\cite{52}.

\begin{table}[htb]
\caption{Total energies (meV) of Ca$_2$$M$ReO$_6$ ($M$=Cr,Fe) with different
magnetization orientations calculated with GGA+SOC, with the lowest energy set as
reference.}\label{table4}
\begin{ruledtabular}
\begin{tabular}{cccccccc}
 $M$ &  [100] & [010] & [001] & [110] & [101] & [011] & [111] \\
\hline
 Cr & 1.00 & 0 & 2.16 & 1.21 & 2.08 & 1.34 & 1.29  \\

 Fe & 1.37 & 0 & 2.04 & 2.03 & 2.62 & 1.60 & 1.61  \\
\end{tabular}
 \end{ruledtabular}
\end{table}


\begin{table}[htb]
\caption{The mBJ results of individual and total spin moments ($\mu^s$), orbital moments
($\mu^o$), and net moments ($\mu_{\rm tot}$) in $\mu_B$, and semiconductor gap ($G_s$, in eV) of Ca$_2$$M$ReO$_6$ ($M$=Cr,Fe) with $P2_1/n$ structure with and without SOC. }\label{table5}
\begin{ruledtabular}
\begin{tabular}{ccccccccc}
$M$ & scheme & $\mu_M^s$ & $\mu_{\rm Re}^s$ &
 $\mu_{\rm tot}^s$ & $\mu_M^o$ & $\mu_{\rm Re}^o$ & $\mu_{\rm tot}$ & $G_{s}$  \\
\hline
 Cr & mBJ      & 2.520 & -1.264 & 1.000 &  &  & 1.000 & 0.38  \\
      & mBJ+SOC  & 2.520 & -1.239 & 1.035 & -0.018 & 0.192 & 1.209 & 0.31 \\ \hline

 Fe & mBJ       & 4.090 & -1.118 & 3.000 & &  & 3.000 & 0.05 \\
      & mBJ+SOC   & 4.090 & -1.096 & 3.033 & 0.042 & 0.183 & 3.258 & 0.03  \\
 \end{tabular}
 \end{ruledtabular}
\end{table}

After the magnetic easy axis is fixed, we do further study with
SOC along the [010] direction. Our calculated spin and orbital moments for the
Ca$_2$$M$ReO$_6$ are summarized in Table \ref{table5}. When SOC is neglected, the total
spin moment is precisely 1$\mu_B$ and 3$\mu_B$ per formula unit for the
Ca$_2$CrReO$_6$ and Ca$_2$FeReO$_6$, respectively. The results can
be elucidated in the ionic model, where the transition-metal
ions are in the ($M$Re)$^{8+}$ valence state. The $M$$^{3+}$ are in the high spin
state of S=3/2 for Cr$^{3+}$ and S=5/2 for Fe$^{3+}$ according to Hund's
rule, antiferromagnetically coupled with highly ionized Re$^{5+}$
with valence spin states of S=1, resulting the integer moment in Bohr magneton.
Note that a large part of the spin moment is delocalized into
the interstitial region, and therefore the spin moments of the
individual $M$ and Re ions appear small compared to the ionic values.

When SOC is taken into account, the total spin moments increase by
0.035$\mu_B$ and 0.033$\mu_B$ for the Ca$_2$$M$ReO$_6$ ($M$=Cr,Fe), respectively,
due to some increase of Re spin moment. The orbital moments of Cr
and Re are both antiparallel to the spin moments, because of the less than
half-filled $d$ shell, in accordance with Hund's rule. As for the Ca$_2$FeReO$_6$,
the orbital moments of Fe $3d$ is of the same sign as the spin
moment, indicating that the $3d$ orbital is more than half-filled, consistent
with Hund's rule. Our calculated Cr orbital moment of -0.018$\mu_B$ is much smaller than the Cr
one of 0.042$\mu_B$, which could be
understood as a consequence of stronger ligand field caused
by the Cr $3d$ orbital than by the Fe $3d$ orbital.
For Re ion, the orbital moment is 0.192$\mu_B$
or 0.183$\mu_B$ for the Ca$_2$CrReO$_6$ or Ca$_2$FeReO$_6$, due to
the strong spin-orbit coupling in $5d$ orbital. The higher improvement
of total magnetic moment, 0.258$\mu_B$,  in the Ca$_2$FeReO$_6$ than 0.209$\mu_B$ in the
Ca$_2$CrReO$_6$ is due to the positive large
orbital moment in Fe ion.

\begin{figure}[htb]
\centerline{\includegraphics[width=7cm]{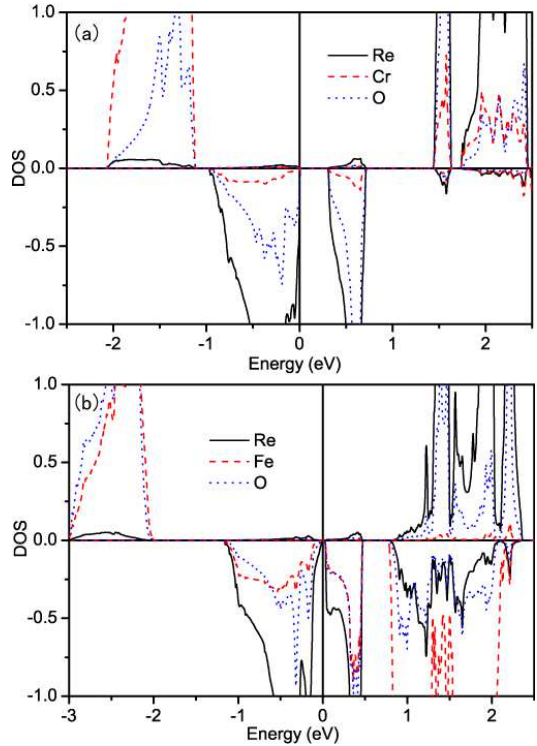}}
\caption{(color online) The spin-resolved DOS
of Ca$_2$CrReO$_6$ (a) and Ca$_2$FeReO$_6$ (b) calculated with mBJ
potential and SOC taken into account.}\label{mBJ+SOC}
\end{figure}

We also present the semiconductor gaps $G_s$ as the
true gaps of these compounds in Table V. When SOC is included, the band gaps remain
but become smaller, and is equivalent to 0.31 eV for the Ca$_2$CrReO$_6$ and 0.03 eV for the Ca$_2$FeReO$_6$,
respectively. In order to understand in more detail the reason of the smaller
gap with SOC, the partial DOS of $M$ and Re projected onto
$d$ orbital in the Ca$_2$$M$ReO$_6$ calculated with SOC are plotted in Fig. 7. In the presence of SOC,
the bands in both of the spin channels hybridize
with each other, in contrast to the pictures in Figs. 3 and 4.
The Re $t_{2g}$ states around the Fermi level in one spin channel
induce some states in the oppositive spin channel, especially in the Ca$_2$FeReO$_6$
compound. Moreover, the Re $t_{2g}$ characteristic states with SOC  are broadened compared
with those without SOC, which leads to the some reduction of the semiconductor gaps of the Ca$_2$$M$ReO$_6$.
It should be pointed out that the noninteger moments in Bohr magneton in the Ca$_2$$M$ReO$_6$ ($M$=Cr,Fe) with $P2_1/n$ structure
are the consequence of the mixing of spin-up and spin-down states,
but the semiconductor gaps are preserved.

\section{Conclusion}

We have used FP-LAPW method to investigate the structural,
electronic, and magnetic properties of double perovskites Ca$_2$$M$ReO$_6$ ($M$=Cr,Fe).
The GGA approach has been used to do geometry optimization, and then the electronic
and magnetic properties have been investigated with mBJ exchange potential for improving on calculational accuracy.
Our calculated results confirm the ground-state $P2_1/n$ structure for both of the Ca$_2$$M$ReO$_6$.
The mBJ calculation reveals the semiconductor characteristic with gaps of
0.38 eV and 0.05 eV, respectively, being consistent with the experimental results
at low temperature\cite{19,16,38,add1,52,33}. The correct mBJ result of the electronic states is because the mBJ potential improves on the DFT description of both crystal field and spin exchange splitting in such compounds.

The origin of semiconductor state of the double perovskite Ca$_2$$M$ReO$_6$ is due to the crystalline distortion of
ReO$_6$ octahedra in the $P2_1/n$ structure, which drives the further splitting of
partially-occupied Re$^{5+}$ $t_{2g}$ state into a full-filled doublet $d_{xy}$+$d_{xz}$ and an empty singlet $d_{yz}$ in the minority-spin channel.
Thus, The semiconductor gap is formed from the crystalline field splitting in one single spin channel, in contrast to the
pervious proposed Mott-Hubbard insulator state. The total spin moments of the
Ca$_2$$M$ReO$_6$ ($M$=Cr,Fe) in the semiconductor ground-states
are found to be 1$\mu_B$ and 3$\mu_B$ per formula unit without the spin-orbit effect included, respectively. When the spin-orbit
coupling is taken into account, the total magnetic moments become noninteger in unit of Bohr magneton as consequence of
the mixing of spin-up and spin-down states. Fortunately, the semiconductor gaps (0.31 and 0.03 eV) remain open
and robust in this structure even when SOC effect is included.

In summary, our DFT investigation with mBJ has established the correct ferrimagnetic semiconductor ground state for the double perovskites Ca$_2$$M$ReO$_6$ ($M$=Cr,Fe), which originate mainly from the crystalline field splitting of the three partially-occupied Re t$_{2g}$ bands into two full-filled doublet bands and one empty singlet band near the Fermi level. This mechanism is different from that in Sr$_2$CrOsO$_6$, and can help understand physical properties of other similar compounds.

\begin{acknowledgments}
This work is supported by Nature Science Foundation of China (Grant
No. 11174359), by Chinese Department of Science and Technology
(Grant No. 2012CB932302), and by the Strategic Priority Research
Program of the Chinese Academy of Sciences (Grant No. XDB07000000).
\end{acknowledgments}

\end{document}